\documentclass[11pt]{article}

\newcommand{\BE}{\begin{equation}}
\newcommand{\EE}{\end{equation}}
\newcommand{\be}{\begin{equation}}
\newcommand{\ee}{\end{equation}}
\newcommand{\BA}{\begin{eqnarray}}
\newcommand{\EA}{\end{eqnarray}}

\usepackage{amssymb}
\usepackage[dvips]{graphicx}

\begin{document}
\begin{titlepage}
\begin{flushright}
{\small DE-FG05-97ER41031-56 }
\end{flushright}
%\vspace*{15mm}
\vspace*{11mm}
\begin{center}
   {\LARGE{\bf How do sound waves}} 
\\
\vspace*{2mm}
   {\LARGE{\bf in a Bose-Einstein condensate }}
\\ 
\vspace*{2mm}
{\LARGE{\bf move so fast?}}

\vspace*{20mm}

{\Large P. M. Stevenson}
\vspace*{3mm}\\
{\large T. W. Bonner Laboratory, Physics Department \\
Rice University, P.O. Box 1892, Houston, TX 77251-1892, USA}
\vspace{32mm}\\
{\bf Abstract:}
\end{center}

     Low-momentum excitations of a dilute Bose-Einstein condensate 
behave as phonons and move at a finite velocity $v_s$.  Yet the atoms 
making up the phonon excitation each move very slowly; $v_a = p/m \to 0$. 
A simple ``cartoon picture'' is suggested to understand this phenomenon 
intuitively.  It implies a relation $v_s/v_a \approx N_{\rm ex}$, where 
$N_{\rm ex}$ is the number of excited atoms making up the phonon.  
This relation does indeed follow from the standard Bogoliubov theory.

\end{titlepage}

\newpage

     In a dilute, non-ideal Bose-Einstein (BE) condensate the spectrum of 
quasiparticle excitations is linear at low momentum: 
$\tilde{E}(p) \approx v_s p$.  Thus, long wavelength excitations are 
{\it phonons} --- quantized sound waves --- that travel at speed $v_s$.  
The standard Bogoliubov theory gives this result and it is also expected on 
general grounds since hydrodynamics applies to any medium at long 
length scales.  

     However, this familiar result has a puzzling aspect to it.  Very 
low momentum atoms move very slowly while very low-momentum phonons move 
at a finite speed $v_s$.  {\it But how can a phonon in a dilute medium move 
so much faster than the atoms that make it up?}  Surprisingly, this issue 
seems not to be addressed in the literature.  The purpose of this brief 
note is to offer an intuitive explanation.  

     It is important to stress that the puzzle arises because we are 
considering a {\it dilute} medium. 
In solids and liquids, where the atoms are ``touching'' (separations 
less than the interaction range), a tiny motion of one 
atom can be transmitted almost immediately to its neighbours, allowing 
compression and other waves to move at speeds much greater than the speed 
of the atoms themselves.  
However, in a dilute gas with only short-range interactions 
(and theoretically the interatomic potential can be replaced with a 
delta-function pseudopotential \cite{huang}) disturbances can only 
be communicated onwards by the atoms themselves travelling from one collision 
to the next.  Thus, the disturbance should propagate no faster than the atoms' 
own speeds.  Indeed, for an ordinary, classical gas the speed of 
sound is of order the average speed (r.m.s. velocity) of the atoms for this 
very reason. 

     Of course, the above intuitive argument is inherently classical.  One 
may immediately observe that it can fail in the BE-condensate case because the 
atoms are not localized entities --- each atom's wavefunction spreads out over 
distances much greater than the interatomic spacing.  While that observation 
is enough to remove any logical paradox it only shows that the trick is 
possible; it does not really explain how the trick is done.  The aim here is 
to offer a more tangible intuitive picture.  (It should be recognized, 
however, that any intuitive explanation of an inherently quantum 
phenomenon will necessarily involve an element of ``quantum mystery'' that 
is not satisfactorily explicable in classical terms.)  

     In the standard theory of a uniform dilute Bose gas \cite{huang,pethick} 
the quasiparticle annihilation operators $b_{\bf k}$ are related to 
the atom annihilation operators $a_{\bf k}$ by a Bogoliubov 
transformation:
\BE
\label{bogtrans}
a_{\bf k} = \frac{1}{\sqrt{1- \alpha_k^2}} 
\left( b_{\bf k} - \alpha_k b_{\bf -k}^\dagger \right) ,
\EE
with 
\BE
\label{alpha}
\alpha_k = 1+ x^2 - x \sqrt{x^2 +2}, \quad \quad \quad 
x^2 \equiv k^2/(8 \pi n a),
\EE
where $n$ is the number density and $a$ is the scattering length.  (We set 
$\hbar = 1$.) 
The quasiparticle spectrum is obtained as 
\BE
\label{specrel}
\tilde{E}(p) = \left( \frac{1+\alpha_p}{1-\alpha_p} \right) E(p)
\EE
where $E(p) = p^2/2m$ is the original atom spectrum.  The resulting 
Bogoliubov spectrum is  
\BE
\label{bogspec}
\tilde{E}(p)= \frac{p}{2m}\sqrt{p^2 + 16 \pi n a } .
\EE
Thus, a low-momentum quasiparticle moves at a finite speed 
$d \tilde{E}/dp \approx \sqrt{4 \pi n a/m^2 } \equiv v_s$.    

      Due to interactions the ground state, even at zero temperature, does 
not have {\it all} the atoms in the ${\bf p} = 0$ mode \cite{huang,pethick}.  
The number of non-condensate atoms is
\BE
N - N_0 = \left\langle \sum_{{\bf k} \neq 0} a^\dagger_{\bf k} a_{\bf k} 
\right\rangle.
\EE
Substituting (\ref{bogtrans}) and using the fact that the ground state is 
annihilated by $b_{\bf k}$ gives 
\BE
\langle 0 |a^\dagger_{\bf k} a_{\bf k} | 0\rangle = \alpha_k^2/(1-\alpha_k^2),
\EE
(which leads to the familiar result for the depletion 
$1-N_0/N = \frac{8}{3} \sqrt{\frac{n a^3}{\pi}}$.)  A similar computation in 
a one-quasiparticle excited state 
$|{\bf p} \rangle \equiv b_{\bf p}^\dagger |0 \rangle $ gives 
\BE
\langle {\bf p} |a^\dagger_{\bf k} a_{\bf k} | {\bf p}\rangle = 
\frac{1}{(1-\alpha_k^2)} \left( \alpha_k^2 + \delta_{{\bf p},{\bf k}} + 
\alpha_p^2 \delta_{-{\bf p},{\bf k}} \right),  
\EE
which shows that an extra number $N_{\rm ex}$ of atoms have been excited 
from the condensate into the ${\bf p}$ and $-{\bf p}$ modes.  Specifically, 
\BE
\label{numbers}
N_{\rm ex} = N_+ + N_- = 
\frac{1}{1-\alpha_p^2} + \frac{\alpha_p^2}{1-\alpha_p^2}.
\EE
Note that $N_+ = N_- + 1$, giving a net momentum ${\bf p}$ to the whole 
excitation.

     When $p$ is large one has $\alpha_p \approx 0$, so that 
$N_{\rm ex} \approx 1$.  Thus, a high-momentum quasiparticle is essentially 
a single moving atom.  For small $p$, however, one has $\alpha_p \to 1$, so 
that $N_{\rm ex}$ becomes very large.  Thus, a low-momentum 
quasiparticle (a phonon) is a coherent state of many atoms with 
momenta ${\bf p}$ and $-{\bf p}$.

       Now we ask: {\it What are all these excited atoms doing?} \ldots 
and how does the collective excitation of slow-moving atoms manage to move 
so fast?  As an answer we suggest the picture in Fig. 1.

\begin{figure}[htp]
\begin{center}
\includegraphics[angle=90, width=12.9cm]{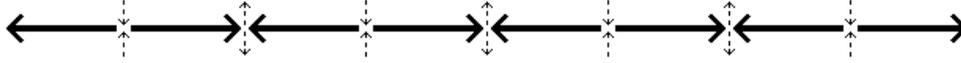}
\caption{Cartoon picture of a phonon.  Solid arrows represent excited atoms 
with momentum ${\bf p}$ or $-{\bf p}$.  Dashed arrows represent condensate 
atoms.}
\end{center}
\end{figure}

This picture shows a series of ``creations'' (where two 
condensate atoms scatter into ${\bf p}$ and $-{\bf p}$ atoms) and 
``annihilations'' (where a ${\bf p}$ and a $-{\bf p}$ atom scatter back 
into two condensate atoms).  (Each ``creation'' event ``borrows'' some energy 
that is repaid in the subsequent ``annihilation'' event.)  The ``creations'' 
occur in rapid succession from left to right, resulting in an apparent 
motion at high speed.  The situation can be seen more clearly in the 
spacetime picture, Fig. 2, in which each individual leg has a steep slope 
(the slope being the inverse of the atom speed), but the overall zig-zag 
path has a shallow slope (corresponding to the high sound speed).

\begin{figure}[htp]
\begin{center}
\includegraphics[angle=90,width=9.5cm]{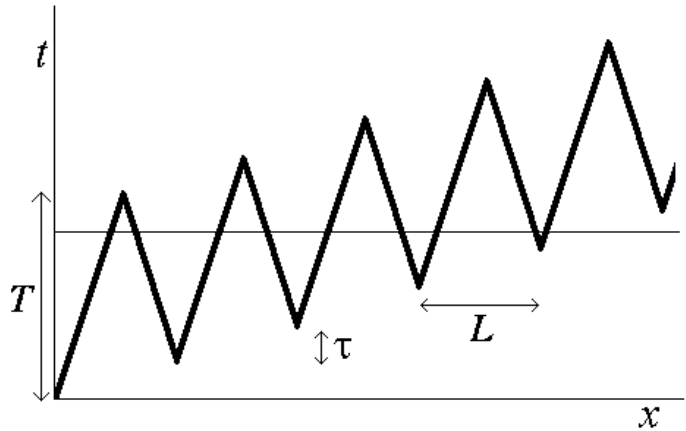}
\caption{Spacetime picture}
\end{center}
\end{figure}

       Of course, this picture is not to be taken too literally since it 
assigns definite trajectories to quantum mechanical particles.  [What we 
have in mind is that path in Fig. 2, together with a set of similar, but 
more irregular zig-zag paths having the same average slope, form the 
class of paths that dominate the multiparticle path integral for the system.]
Note that the arrangement of ``creation'' and annihilation'' events in quick 
succession requires a sort of ``conspiracy'' --- but one that is possible 
given the quantum coherence of the condensate as a whole.  (This 
``conspiracy'' represents the element of ``quantum mystery'' referred to 
earlier as unavoidable in any attempt to intuitively explain an inherently 
quantum-mechanical phenomenon.)

       Let us examine some consequences of the picture in Fig. 2 (taking it 
literally, despite our previous caveat).  At any given instant, 
corresponding to the thin horizontal line, there are many excited atoms 
present; half left-moving and half (plus one extra) right-moving.  If 
neighbouring creation events are separated in space by $L$ and separated in 
time by $\tau$ then the overall speed of the zig-zag path --- the 
sound speed --- is $v_s = L/\tau$.  Let the time interval between a 
``creation'' and a subsequent ``annihilation'' be $T$ (with $T \gg \tau$ 
when $N_{\rm ex} \gg 1$).  Each atom covers a distance $L/2$ in time $T$, 
giving an atom speed of $v_a = \frac{L}{2 T}$.  Hence the ratio 
$\frac{v_s}{v_a} = \frac{L}{\tau}/\frac{L}{2 T} = \frac{2 T}{\tau}  
\approx N_{\rm ex}$.  That is, the spacetime picture implies, for low 
momentum, the formula
\BE
           v_s/v_a = N_{\rm ex}.
\EE  

        Since the velocities follow from differentiating the respective 
spectra, and since $N_{\rm ex}$ is given by (\ref{numbers}), the 
above equation is 
\BE
\label{dedp}
\frac{d \tilde{E}(p)}{dp} = 
\left( \frac{1+\alpha_p^2}{1-\alpha_p^2} \right) \frac{dE}{dp}.
\EE
This relationship indeed follows from (\ref{specrel}) after some algebra, 
employing the properties of $\alpha_p$ of (\ref{alpha}).  In fact 
Eq. (\ref{dedp}) is true at all $p$, although the connection with our 
``cartoon picture'' is valid only for small $p$.  
  
    The preceding considerations do not fix the absolute length scale.  
However, since Fig. 1 does indeed look like a density undulation with a 
wavelength $L$ we would expect $L$ to be $2 \pi/p$ for momentum (wavenumber) 
$p$.  We can justify this relation, in order-of-magnitude terms, as follows.  
Each ``creation'' event ``borrows'' an energy $2 E(p) = p^2/m$, which is 
only ``repaid'' at the subsequent neighbouring ``annihilations'' a time $T$ 
later.  In the usual way this violation of local energy conservation can 
hide under the uncertainty principle provided $T \approx 1/(2 E)$.  This in 
turn implies that $L \approx 2 T v_a \approx \frac{2m}{p^2} \frac{p}{m}$, 
and so $L$ is of order $1/p$, as expected.   

    Experiments, notably by Ketterle's group \cite{sound,bogtrans}, have 
directly verified sound propagation in BE condensates.  In such atom-trap 
experiments the longest wavelengths accessible are limited by the 
diameter of the cloud, hence $p_{\rm min} \approx \hbar \pi/R$, where $R$ 
is the Thomas-Fermi radius of the cloud in the appropriate direction.  This 
consideration yields an upper limit 
\BE
       \frac{v_s}{v_a} \lesssim 0.22 \, \frac{R}{\xi}
\EE
where $\xi\equiv 1/\sqrt{8\pi n a}$ is the healing length, related to 
$v_s$ by $v_s = \hbar/(\sqrt{2} m \xi)$.  The ratio of length scales may 
may also be written as \cite{baym,pethick} 
\BE
\frac{R}{\xi} = \frac{R^2}{a_{\rm osc}^2} = 
\frac{a_{\rm osc}^2}{\xi^2},
\EE
where $a_{\rm osc} = \sqrt{\hbar/m \omega}$ is the nominal trap size, set 
by the angular frequency $\omega$ of the harmonic trapping potential.   

     For a typical $^{87}{\rm Rb}$ trap with central density 
$n=10^{15}~{\rm cm}^{-3}$, $\omega/2 \pi = 150~{\rm Hz}$, and 
$a \sim 5.3~{\rm nm}$ one finds that $v_s/v_a \lesssim 20$.  For the 
long $^{23}{\rm Na}$ trap of Ref. \cite{sound} with $2R \sim 450~\mu{\rm m}$ 
and $\xi \sim 0.2~\mu{\rm m}$, the maximum velocity ratio is about $250$.  
However, since the length of the pulses created was only about 
$20~\mu{\rm m}$, the actual $v_s/v_a$ ratio for the phonon states studied 
was around $20$.  That is still well into the region where the ``puzzle'' 
of $v_s \gg v_a$ is an issue.     

\begin{center}
{\bf Acknowledgements}
\end{center}

I thank Randy Hulet for helpful comments.  This work was supported in part 
by the Department of Energy under Grant No. DE-FG05-97ER41031.

%\newpage

\end{document}